\documentclass[aps,prl,reprint,amsmath,amssymb,showpacs,superscriptaddress]{revtex4-1}
\usepackage{graphicx}% Include figure files
\usepackage{dcolumn}% Align table columns on decimal point
\usepackage{bm}% bold math

\usepackage{hyperref}% add hypertext capabilities
\begin{document}
%\begin{CJK*}{UTF8}{} % Use default fonts from CJK (see below)


\title{Radii of neutron drops probed via the neutron skin thickness of nuclei}

\author{P. W. Zhao %({\CJKfamily{gbsn}} 赵鹏巍)
}
%\email{pwzhao@pku.edu.cn}
%\thanks{}
\affiliation{Physics Division, Argonne National Laboratory, Argonne, Illinois 60439, USA}

\author{S. Gandolfi 
}
\affiliation{Theoretical Division, Los Alamos National Laboratory, Los Alamos, New Mexico 87545, USA}

%\date{\today}

\begin{abstract}
Multi-neutron systems are crucial to understanding the physics of neutron-rich
nuclei and neutron stars.  Neutron drops, neutrons confined in an external field, 
are investigated systematically in both non-relativistic and relativistic 
density functional theories and with {\it ab initio} calculations.
We demonstrate a new strong linear correlation, which is universal in the
realm of mean-field models, between the rms radii of
neutron drops and the neutron skin thickness of  $^{208}$Pb and $^{48}$Ca; i.e., the difference
between the neutron and proton rms radii of a nucleus.  Due to its high
quality, this correlation can be used to deduce the radii of neutron drops
from the measured neutron skin thickness in a model-independent way,
and the radii obtained for neutron drops can provide a useful constraint
for realistic three neutron forces. 
We also present a new correlation between the slope $L$ of the symmetry energy and the 
radii of neutron drops, and provide the first validation of such a correlation
by using density-functional models and {\it ab initio} calculations.
These newly established correlations, together with more precise measurements of the neutron skin thicknesses of $^{208}$Pb
and $^{48}$Ca and/or accurate determinations of $L$, will have an enduring impact on the understanding of
multi-neutron interactions, neutron-rich nuclei, neutron stars, etc.

\end{abstract}

% 21.60.Jz Nuclear Density Functional Theory and extensions 
%21.10.Gv	Nucleon distributions and halo features
%21.30.-x	Nuclear forces
%21.10.-k Properties of nuclei; nuclear energy levels
%21.10.Re Collective levels
%21.60.Ev Collective models
%21.60.Cs	Shell model
%21.45.Ff	Three-nucleon forces
%23.20.-g Electromagnetic transitions
%23.20.Js Multipole matrix element
%27.20.+n  6 A 19
%27.60.+j 90  A 149
%27.50.+e 59  A  89
%\keywords{}

\maketitle

%\end{CJK*}


%\section{Introduction}
%=======================================================================================
Pure neutron systems have attracted considerable attention in nuclear
physics, since their properties are crucial for understanding
neutron-rich systems ranging from microscopic rare isotopes at
the femtometer scale to macroscopic neutron stars.
On the one hand, they are very useful to
probe possible new physics for nuclei with large isospin in as yet
unexplored regions of the nuclear chart~\cite{Erler2012Nature509}.
Such interests are supported further by the advent of new
rare-isotope facilities~\cite{Tanihata1985Phys.Rev.Lett.2676}
and also by the quest to understand the origin of
the elements in the Universe through nucleosynthesis
processes~\cite{Qian2003Prog.Part.Nucl.Phys.153,Arnould2007Phys.Rep.97}.
On the other hand, understanding the inner crust of neutron
stars also requires accurate knowledge of inhomogeneous neutron
matter~\cite{Ravenhall1983Phys.Rev.Lett.2066,Brown2009Astrophys.J.1020,Buraczynski2016Phys.Rev.Lett.152501}.

Although a candidate resonant tetraneutron state was proposed
recently~\cite{Kisamori2016Phys.Rev.Lett.52501}, most multi-neutron
systems are not self-bound and, thus, an external potential must
be employed to produce bound states; i.e., the so-called ``neutron
drops''.  Due to its simplicity,
a neutron drop can serve as a unique test case for various nuclear
many-body methods; e.g., {\it ab initio} approaches for light nuclei
and density-functional theories (DFTs) for heavy ones.  The former solve
directly a many-body Hamiltonian with realistic nucleon-nucleon ($NN$) and
three-nucleon (3$N$) interactions, while the latter resort to a variation
of an energy functional with respect to nucleon densities.
Moreover, neutron drops also provide an essential test for density
matrix expansion (DME) techniques, which aim to build DFTs from realistic
$NN$ and 3$N$ interactions~\cite{Bogner2011Phys.Rev.C44306}, and can describe properties
of neutron-rich nuclei~\cite{Gandolfi2006Phys.Rev.C44304,Gandolfi2008Eur.Phys.J.A207}.

So far, neutron drops have been studied with many {\it
ab initio} approaches.  Quantum Monte Carlo
(QMC)~\cite{Carlson2015Rev.Mod.Phys.1067} studies for neutron drops can
be traced back to the 1990s, and only light droplets with $N=6,7,8$
were calculated at that time~\cite{Pudliner1996Phys.Rev.Lett.2416}.
Systematic calculations, covering a wide range of neutron
numbers and external potentials, have been performed recently with high-accuracy phenomenological and chiral $NN+3N$ interactions~\cite{Gandolfi2011Phys.Rev.Lett.12501,Maris2013Phys.Rev.C54318,Potter2014Phys.Lett.B445,Tews:2016}.
These {\it ab initio} solutions for neutron drops provide important
references for nuclear energy density functionals which are
usually determined by fitting to available nuclei.  In comparison
with the QMC calculations, traditional Skyrme density functionals
considerably overbind neutron drops and yield too-large a spin-orbit
splitting~\cite{Pudliner1996Phys.Rev.Lett.2416,Gandolfi2011Phys.Rev.Lett.12501}.
%Such significant differences have been removed in recent UNEDF
%functionals~\cite{Kortelainen2014Phys.Rev.C54314}.

All density-functional studies of neutron
drops hitherto are in the framework of non-relativistic DFTs, and a
relativistic study is still missing.  Relativistic (covariant) DFTs,
which invoke a different organization of the nuclear many-body problem,
are particularly compelling because the spin degrees of freedom dictated
by relativity can be naturally included~\cite{Lalazissis2004,Meng2015}.
On the other hand, the existing predictions vary largely among different
theories, since neither the isospin $T=3/2$ component of the 3$N$
force~\cite{Pieper2003Phys.Rev.Lett.252501,Hammer2013Rev.Mod.Phys.197}
nor the isovector parts (depending on the difference of the neutron and proton densities) 
of the density functionals~\cite{Reinhard2010Phys.Rev.C51303}
are clearly known.

Since it is not possible to carry out direct
measurements, a connection between
the neutron drop and an isospin-sensitive observable in finite nuclei
can be of help in further understanding the properties of neutron drops.
The neutron skin thickness; i.e., the difference between neutron
and proton rms radii $\Delta r_{np} = r_n-r_p$,
is a typical isospin-sensitive observable for finite nuclei.  Its
connection to the symmetry energy of nuclear matter has attracted a lot of
attention~\cite{AlexBrown2000Phys.Rev.Lett.5296,Centelles2009Phys.Rev.Lett.122502,Tsang2012Phys.Rev.C15803,Horowitz2014J.Phys.G93001}.
Moreover, worldwide efforts have been made to measure
$\Delta r_{np}$ through parity-violating electron
scattering~\cite{Abrahamyan2012Phys.Rev.Lett.112502}, coherent pion
photoproduction~\cite{Tarbert2014Phys.Rev.Lett.242502}, elastic
proton scattering~\cite{Zenihiro2010Phys.Rev.C44611}, antiprotonic
atoms~\cite{Trzci'nska2001Phys.Rev.Lett.82501,Klos2007Phys.Rev.C14311},
electric dipole
polarizability~\cite{Tamii2011Phys.Rev.Lett.62502,Rossi2013Phys.Rev.Lett.242503},
and other methods.  In particular, by measuring parity-violation in electron scattering,
the lead and calcium radius experiments, PREX and CREX, at the Thomas Jefferson National Accelerator Facility (JLab) aim to
provide a $\Delta r_{np}$ value for $^{208}$Pb and $^{48}$Ca independent of most
strong interaction uncertainties~\cite{Horowitz2001Phys.Rev.C25501}.

In this Rapid Communication, we focus on the connection between the observable
$\Delta r_{np}$ and neutron drops.  To this end, a systematic
investigation of neutron drops has been carried out with both
non-relativistic and relativistic
DFTs.  Many well-determined density
functionals, widely used for nuclear and astrophysical problems,
have been employed in the calculations. By doing so, a strong linear
correlation between the neutron skin thickness $\Delta r_{np}$  and
the rms radii $R$ of neutron drops is revealed.  This correlation is
universal with respect to the variation in neutron number and the
strength of an external field for neutron drops, once the central density
of the drop is close to nuclear saturation density.  This allows one to
extract $R$ of neutron drops from the measured neutron skin thickness
in a model-independent way.

The skin thickness of large nuclei is correlated to 
the slope $L$ of the symmetry energy, which is directly related to the 
equation of state (EOS) of pure neutron matter and to the radii of 
neutron stars~\cite{Gandolfi2012Phys.Rev.C32801,Steiner2012Phys.Rev.Lett.81102,Fattoyev2012Phys.Rev.C15802}.
In particular, a strong linear correlation between 
the skin thickness of $^{208}$Pb and the value of $L$ has been demonstrated within mean-field models~\cite{Roca-Maza2011Phys.Rev.Lett.252501}. 
In this Rapid Communication, we will also discuss the relation between 
radii of neutron drops and the value of $L$ from the EOS 
in the framework of both mean-field models and {\it ab initio} calculations with several microscopic Hamiltonians. 
We employ the auxiliary field diffusion Monte Carlo (AFDMC) 
method~\cite{Schmidt1999Phys.Lett.B99} to calculate the energy and radii of 20 neutrons in 
a harmonic oscillator trap~\cite{Gandolfi2011Phys.Rev.Lett.12501} and
the EOS of neutron matter~\cite{Gandolfi2009Phys.Rev.C54005} using several nuclear Hamiltonians,
including the Argonne AV8$^\prime$ and AV8$^\prime$+UIX~\cite{Pudliner1997Phys.Rev.C1720},
and local chiral forces at next-to-next-to-leading-order (N$^2$LO)~\cite{Gezerlis2013Phys.Rev.Lett.32501,Gezerlis2014Phys.Rev.C54323}.

%\section{Theoretical Framework}
%-----------------------------------------------------
We first present the first relativistic study of neutron drops in the
framework of covariant DFT. The approach starts from a Lagrangian and
the corresponding Kohn-Sham equations have the form of a Dirac equation
with effective fields $S(\bm{r})$ and $V(\bm{r})$ derived from this
Lagrangian~\cite{Ring1996Prog.Part.Nucl.Phys.193}.  For neutron drops,
these fields are assumed to be spherical and the calculations are carried
out with an external field $V_{ex}$ which, in this work, has the form of
an harmonic oscillator (HO) field, 
\begin{equation}
\label{Diracequation}
   [\bm{\alpha}\cdot\bm{p}+\beta(m+S)+V+V_{ex}]\psi_k=\epsilon_k\psi_k.
\end{equation}
Here $V_{ex} = (m\omega^2/2)r^2$ is the external HO field with
$\hbar^2/m=41.44~\rm {MeV~fm^2}$.  The fields $S$ and $V$ are
connected in a self-consistent way to densities, so this equation
requires an iterative solution, which yields the total energies, rms
radii, etc.  The pairing correlations are considered by solving the
full relativistic Hartree-Bogoliubov (RHB) problem with a separable
pairing force~\cite{Tian2009Phys.Lett.B44}.  For details, see
Refs.~\cite{Ring1996Prog.Part.Nucl.Phys.193,Vretenar2005Phys.Rep.101,Meng2006Prog.Part.Nucl.Phys.470}.

%\section{Results and discussion}
%-----------------------------------------------------
\begin{figure}[!htbp]
\centering
\includegraphics[width=8.5cm]{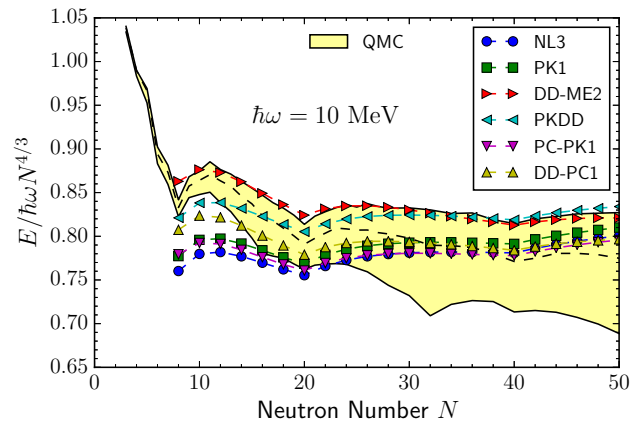}
\caption{(color online) Total energies (scaled by $\hbar\omega N^{4/3}$)
of $N$-neutron systems in a HO trap ($\hbar\omega =
10$ MeV) predicted with various relativistic density functionals (solid
symbols). The shaded area indicates the {\it ab initio} QMC results from
Refs.~\cite{Gandolfi2011Phys.Rev.Lett.12501,Maris2013Phys.Rev.C54318}
for the $NN$ interaction AV8$^\prime$ (center dashed line) as well as
the AV8$^\prime$ plus 3$N$ interactions of UIX (upper solid line) and IL7
(lower solid line).}
\label{fig1}
\end{figure}

The calculations were carried out with six typical relativistic
density functionals shown to be  successful in many applications
for finite nuclei.  They cover nearly all the existing ways
to build a relativistic functional; i.e., the nonlinear meson
exchange functionals NL3~\cite{Lalazissis1997Phys.Rev.C540}
and PK1~\cite{Long2004Phys.Rev.C34319}, density-dependent meson
exchange ones DD-ME2~\cite{Lalazissis2005Phys.Rev.C024312} and
PKDD~\cite{Long2004Phys.Rev.C34319}, a nonlinear point coupling one
PC-PK1~\cite{Zhao2010Phys.Rev.C54319}, as well as the density-dependent
point coupling one DD-PC1~\cite{Niksic2008Phys.Rev.C34318}.
In Fig.~\ref{fig1}, the calculated total energies for neutron drops
are presented by scaling with the Thomas-Fermi $N$-dependence
($N^{4/3}$) and the HO strength ($\hbar\omega = 10$ MeV).  For
comparison, the {\it ab initio} QMC results obtained in
Refs.~\cite{Gandolfi2011Phys.Rev.Lett.12501,Maris2013Phys.Rev.C54318},
are also given, where the AV8$^\prime$ $NN$ interaction~\cite{Pudliner1997Phys.Rev.C1720}
is used with two different 3$N$ interactions, Urbana IX
(UIX)~\cite{Pudliner1997Phys.Rev.C1720} and Illinois-7
(IL7)~\cite{Pieper2008AIPConf.Proc.143}.

The 3$N$ force IL7 is known to be far too attractive at high 
densities in pure neutron systems~\cite{Sarsa:2003,Maris2013Phys.Rev.C54318}, and
it conflicts with the observations in two-solar-mass neutron stars.
Thus, as indicated in Fig.~\ref{fig1}, the
significant reduction of energies given by adding IL7 to AV8$^\prime$
for the droplets with large neutron numbers should not be viewed
as reliable.  All the
density-functional results here are larger than those given solely
by the AV8$^\prime$ Hamiltonian at large neutron numbers.  This is
also consistent with the results given by adding the UIX interaction
to AV8$^\prime$, in particular, for the PKDD and DD-ME2 functionals.
Although other functionals provide slightly lower energies, it should
be noted that the recent {\it ab initio} study with chiral Hamiltonians
indicates only weak contributions from the inclusion of the chiral 3$N$
forces~\cite{Potter2014Phys.Lett.B445}.

For light nuclei (up to $A$=12), however, IL7 provides a much better
description than either AV8$^\prime$ or AV8$^\prime+$UIX, which
typically underbind these nuclei~\cite{Carlson2015Rev.Mod.Phys.1067}.  
This suggests that the IL7 force
may be more reliable for the droplets with small neutron numbers.
Moreover, Ref.~\cite{Maris2013Phys.Rev.C54318} demonstrates that
the AV8$^\prime+$IL7 results below 12 neutrons are, indeed,
very similar to the no-core-shell-model (NCSM) ones with a nonlocal $NN$ interaction
JISP16~\cite{Shirokov2007Phys.Lett.B33}, which also gives a good
description of light nuclei.  The density-functional results in
Fig.~\ref{fig1}, except those for the DD-ME2 functional, are also closer
to AV8$^\prime+$IL7 at small neutron numbers, in particular those computed
with the PKDD and DD-PC1 functionals.
Figure~\ref{fig1} also shows that the uncertainties associated with the
realistic 3$N$ forces and the isovector parts of the density functionals
are almost at the same level when predicting neutron drop properties.
Therefore, an experimental knowledge can be very helpful to probe both.
However, since it is hardly possible to directly measure a neutron
drop, we intend here to connect
its properties to an isospin-sensitive observable of finite nuclei;
e.g., the neutron skin thickness.

\begin{figure}[!htbp]
\centering
\includegraphics[width=8.5cm]{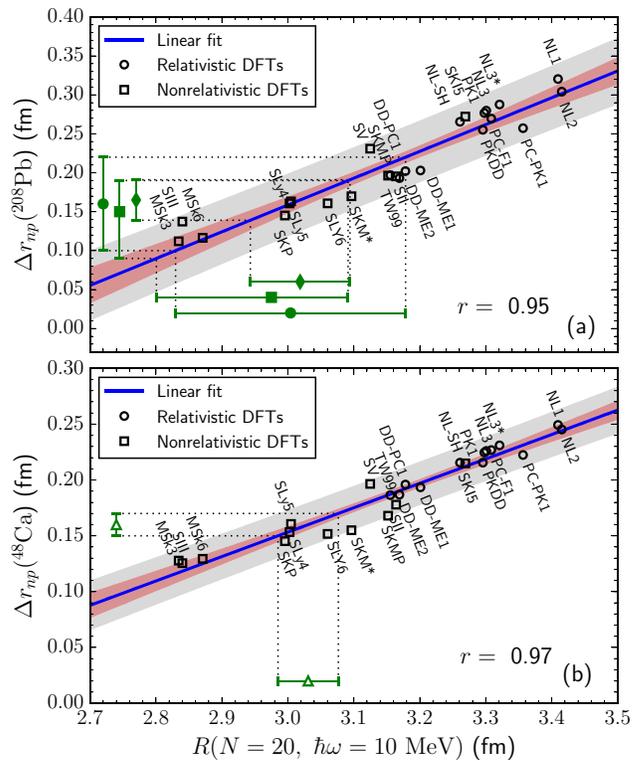}
\caption{(color online) (a) Neutron skin thickness $\Delta
r_{np}$ of $^{208}$Pb against the rms radius $R$ of 20 neutrons
trapped in a HO potential with $\hbar\omega =$ 10 MeV for various
nuclear density functionals. The inner (outer) colored regions
depict the 95\% confidence (prediction) intervals of the linear
regression, and Pearson's correlation coefficient $r$
is also displayed (the $r$ is a value between $+1$ and $−1$ inclusive, 
where $+1$ is total positive correlation, $0$ is no correlation, 
and $−1$ is total negative correlation; see, e.g., Ref.~\cite{Draper1998}). 
The data of $\Delta r_{np}$ in different measurements with
antiprotonic atoms~\cite{Klos2007Phys.Rev.C14311} (circle), pion
photoproduction~\cite{Tarbert2014Phys.Rev.Lett.242502} (square), and
electric dipole polarizability~\cite{Roca-Maza2013Phys.Rev.C24316}
(diamond), are also given together with their projections on the radius
of the neutron drop.
(b) Same plot, but for the neutron skin thickness $\Delta r_{np}$ of $^{48}$Ca.
The estimate of $\Delta r_{np}$ is from a prediction of electric dipole
polarizability~\cite{Roca-Maza2015Phys.Rev.C64304}. 
}
\label{fig2} 
\end{figure}

We computed the neutron skin thicknesses $\Delta r_{np}$ of $^{208}$Pb and $^{48}$Ca 
using a large sample of nuclear density functionals
based on very different schemes: from non-relativistic to relativistic
ones~\cite{Bender2003Rev.Mod.Phys.121}, from finite range meson-exchange
to zero-range point-coupling ones~\cite{Meng2015}.  Figure~\ref{fig2}
depicts $\Delta r_{np}$ obtained for $^{208}$Pb and $^{48}$Ca as a function of the rms
radius $R$ of 20 neutrons trapped in a HO potential with $\hbar\omega =$
10 MeV.  All the functionals considered here are quite successful in
describing bulk properties such as binding energies and charge radii
for nuclei over the entire nuclide chart.  However, one can clearly
see in Fig.~\ref{fig2} that their predictions for $\Delta r_{np}$
are very different, from 0.1 to 0.3 fm, since the isovector channels
in these phenomenological functionals are loosely determined in the
fitting procedures.  The functionals with softer (stiffer) symmetry
energy at the saturation density yield smaller (larger) $\Delta r_{np}$
values~\cite{Centelles2009Phys.Rev.Lett.122502}.

A strong linear correlation is found between the neutron skin
thickness $\Delta r_{np}$ and the rms radius $R$
of the 20 neutrons in the potential with $\hbar\omega = 10$ MeV.
The Pearson's correlation coefficient is $r = 0.95$ (see, e.g.,
Ref.~\cite{Draper1998}) for $^{208}$Pb and $r=0.97$ for $^{48}$Ca.  
We note that this strong linear correlation
is universal in the realm of mean-field theory, since it is based on
widely different nuclear density functionals.  It reflects the fact
that both the neutron skin thickness and the radius of a neutron
drop are highly relevant to the behavior of the nuclear symmetry
energy.  Such a high quality linear correlation allows one to deduce
the rms radius $R$ from the measured $\Delta r_{np}$.
In Fig.~\ref{fig2}(a), the measured $\Delta r_{np}$ of $^{208}$Pb from antiprotonic
atoms (circle)~\cite{Klos2007Phys.Rev.C14311}, pion photoproduction
(square)~\cite{Tarbert2014Phys.Rev.Lett.242502}, and electric dipole
polarizability (diamond)~\cite{Roca-Maza2013Phys.Rev.C24316} are shown.
These data have their central values around 0.15 fm and agree well
with each other within the errors.  They determine, through the linear
fit of Fig.~\ref{fig2}(a), that the rms radius $R$ has a central value
around 3.0 fm.  Note that the 16\% accuracy in $\Delta r_{np}$ from the
electric dipole polarizability leads to a $\sim$2.5\% accuracy in $R$
of the neutron drop.
There is also a large set of experiments which suggests a larger neutron skin $\Delta r_{np}\sim0.2$ fm for $^{208}$Pb (see Table~1 in Ref.~\cite{Krasznahorkay2013PhysicaScripta14018}). In particular, the first PREX experiment gives $\Delta r_{np} = 0.33^{+0.16}_{-0.18}$ fm~\cite{Abrahamyan2012Phys.Rev.Lett.112502}. 
This value is not shown in Fig.~\ref{fig2} due to its large error bar. However, the usefulness of the correlation described here can be easily repeated, once we know the result from the upcoming PREX-II experiment, which is aimed at reducing the uncertainty by a factor of 3~\cite{PREX-II}. 

Apart from $^{208}$Pb, the neutron skin thickness $\Delta r_{np}$ of
$^{48}$Ca has been another recent focus of experiment,
and it may provide key information for bridging DFTs and {\it ab initio}
approaches~\cite{Hagen2016NaturePhys186}. Since the
experiments for $\Delta r_{np}$ of $^{48}$Ca are still ongoing, an
estimate of $\Delta r_{np} = 0.16\pm0.01$~fm from a prediction of
electric dipole polarizability~\cite{Roca-Maza2015Phys.Rev.C64304} is
shown in Fig.~\ref{fig2}(b).  The resulting $R$ for the neutron drop from
the linear fit is $R = 3.04\pm0.04$ fm, which is consistent with the $R$
value determined from the $\Delta r_{np}$ of $^{208}$Pb.  Note that, in
this case, the 6.25\% accuracy in $\Delta r_{np}$ leads to an accuracy
of 1.3\% in $R$ for the neutron drop.

\begin{figure}[!htbp]
\centering
\includegraphics[width=8.5cm]{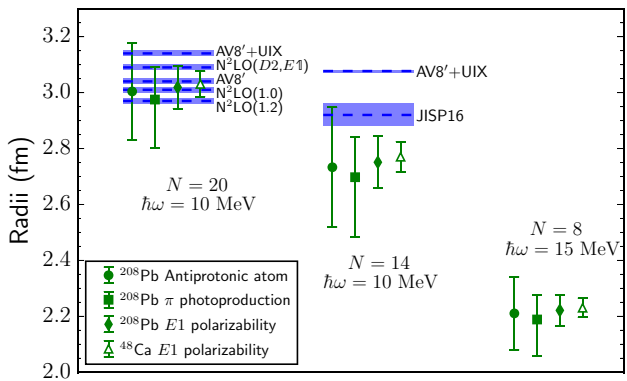}
\caption{(color online) The rms radii for three neutron drops determined
from their linear correlations with the neutron skin thicknesses
of $^{208}$Pb and $^{48}$Ca. For comparison, the {\it ab initio} results obtained using 
phenomenological forces and local chiral forces of Ref.~\cite{Lynn2016Phys.Rev.Lett.62501} 
are also shown.}
\label{fig3} 
\end{figure}

There is no particular reason to study 20 neutrons, other than the fact that it is a closed shell configuration. 
In fact, the linear correlations found for $N=20$ are also obtained for neutron
drops with different neutron numbers and different external traps.   
The only requirement inherent
in this result is that the central density of the neutron drop does not
differ greatly from the saturation density ($\sim 0.16~{\rm fm}^{-3}$).
This condition can be readily satisfied by varying the strength of
the external field; i.e., larger (smaller) neutron numbers should be
associated with weaker (stronger) external fields.
We have considered three different neutron drop systems, with 20 and 14 
neutrons in a HO with $\hbar\omega=10$ MeV, and 8 neutrons in 15 MeV. 
For all three systems, strong linear correlations between their rms radii and the
neutron skin thickness $\Delta r_{np}$ of $^{208}$Pb or $^{48}$Ca
are found. The resulting rms radii from different data of $\Delta
r_{np}$ are compared in Fig.~\ref{fig3} with the 
AFDMC calculations for 20 neutrons obtained using different nuclear Hamiltonians including the Argonne AV8$^\prime$ and AV8$^\prime$+UIX~\cite{Gandolfi2011Phys.Rev.Lett.12501}, and local chiral forces of Ref.~\cite{Lynn2016Phys.Rev.Lett.62501}. 
We have also calculated the radial density, 
and verified that, in the center, the density of the drop is always 
within $0.16\pm0.02~{\rm fm}^{-3}$. 

The determined radii of the $N=14$ droplet from the skin thickness are smaller than 
the results with AV8$^\prime$+UIX~\cite{Gandolfi2011Phys.Rev.Lett.12501}, showing that the UIX force might be too repulsive here. 
For the $N=20$ droplet, the radii obtained from AFDMC with local chiral forces are smaller than those with AV8$^\prime$ and UIX forces, and they agree quite well with the radii determined by the skin thicknesses.
With the development of the high-accuracy measurements of neutron skin thickness, especially the PREX and CREX programs at JLab~\cite{PREX-II}, the radii of neutron drops will be deduced more accurately in the near future. 

\begin{figure}[!htbp]
\centering
\includegraphics[width=8.5cm]{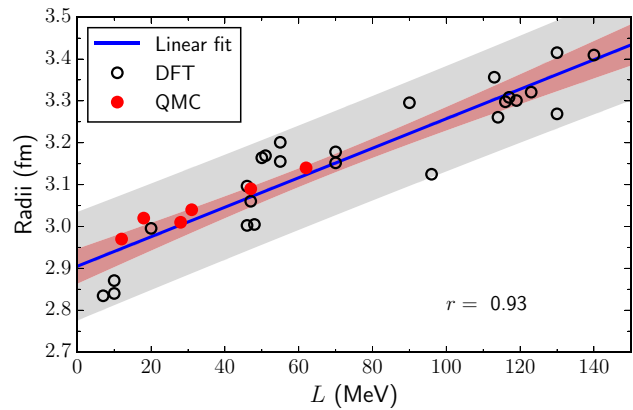}
\caption{(color online) The values of the slope $L$ of the symmetry energy against rms radii of 20 neutrons in a HO with $\hbar\omega=10$ MeV, obtained with density functional theories (open circles) and {\it ab initio} methods (solid circles).} 
\label{fig4} 
\end{figure}

Finally, we have calculated the EOS of pure neutron matter using AFDMC
to fit the slope $L$ of the symmetry energy as discussed in Refs.~\cite{Gandolfi2012Phys.Rev.C32801,Gandolfi2014Eur.Phys.J.A10}.
Since the radius $R$ of neutron drops is correlated with the skin thickness of nuclei,
and the skin thickness with the value of $L$~\cite{Roca-Maza2011Phys.Rev.Lett.252501}, 
it is interesting to plot these quantities together, see Fig.~\ref{fig4}. 
The values of $L$ for the various
functionals considered are taken from Refs.~\cite{Dutra2012Phys.Rev.C35201,Dutra2014Phys.Rev.C55203}.
We find that the two quantities $R$ and $L$ are a bit less well correlated than $R$ vs 
$\Delta r_{np}$, or $\Delta r_{np}$ vs $L$ of Ref.~\cite{Roca-Maza2011Phys.Rev.Lett.252501}. 
The Pearson's coefficient is obtained as $r=0.92$ by fitting the density-functional results.
It is interesting to note that the density-functional predictions of the correlation between $R$ and $L$ is compatible with  {\it ab initio} calculations.
This provides the first validation of such a new correlation between $L$ and pure neutron systems with {\it ab initio} calculations.
The slope $L$ has been related to many other nuclear properties and, thus, can be determined by various ways, though with currently large uncertainties.
Therefore, we note that important constraints on the three-neutron force can be obtained when $L$ or the neutron skin thickness of nuclei are accurately measured. 

%\section{Summary}
%-----------------------------------------------------
In summary, the properties of neutron drops have been investigated systematically with DFTs 
in comparison with results from {\it ab initio}
calculations. In particular, this is the first investigation of neutron drops using relativistic functionals. 
The uncertainties of the realistic 3$N$ forces and the
isovector parts of density functionals are found to be large and
comparable for predicting neutron drop properties.  A new strong linear
correlation between the rms radii of neutron drops and the neutron
skin thicknesses of $^{208}$Pb and $^{48}$Ca has been demonstrated.
This correlation is universal in the realm of density functional theories,
and applies to different neutron drops. 
% if their central densities do not significantly differ from saturation density. 
Due to its
high quality, this linear correlation can be used to deduce the radii of
neutron drops by measuring the neutron skin thickness, and these radii
can in turn provide a useful constraint for realistic 3$N$ forces. 
In view of upcoming high-precision measurements of neutron skin thicknesses in
$^{208}$Pb and $^{48}$Ca, this correlation is likely to have an enduring
impact on the understanding of multi-neutron interactions.
We have also provided the first validation of a new correlation between 
radii of neutron drops with the slope of the symmetry energy by using density-functional 
models and {\it ab initio} calculations. 
In this case, the density-functional results are very close to the {\it ab initio} ones, suggesting
that radii of confined neutrons can give important information of the 
slope of the symmetry energy. Future similar calculations of radii of neutron 
drops in different external traps might open the way to calculating and predicting
 $L$ at different densities.

\begin{acknowledgments}
The authors thank Steven C. Pieper, Robert B. Wiringa, and R. V. F. Janssens for valuable
discussions and careful reading of the manuscript.
This work is supported by U.S. Department
of Energy (DOE), Office of Science, Office of Nuclear Physics, under
Contracts No. DE-AC02-06CH11357 (P.W.Z.) and No. DE-AC52-06NA25396 (S.G.)
and by the NUCLEI SciDAC project (S.G.).
Computational resources have been provided by the
Laboratory Computing Resource Center at Argonne National Laboratory
and by Los Alamos Open Supercomputing.
This research used resources of the National Energy Research
Scientific Computing Center, a DOE Office of Science User Facility 
supported by the Office of Science of the U.S. Department of Energy 
under Contract No. DE-AC02-05CH11231.
\end{acknowledgments}

%\bibliography{paper}


\end{document}